\documentclass[aps,prd,10pt,preprint,superscriptaddress,onecolumn,nofootinbib]{revtex4-1}

\usepackage{graphicx, epsfig} 
\usepackage{amsmath,amssymb,amsfonts,dsfont,mathrsfs,amsthm,mathtools}
\usepackage{bm} 
\usepackage{color}
\usepackage[usenames]{xcolor}
\usepackage{hyperref}
\usepackage{siunitx}
\hypersetup{colorlinks=true,urlcolor=blue,linkcolor=magenta,citecolor=blue,filecolor=blue}
\usepackage[normalem]{ulem}
\usepackage{array}
\usepackage{hyperref}
\usepackage{booktabs}
\usepackage{blindtext}

\definecolor{armygreen}{rgb}{0.29, 0.33, 0.13}

\newcommand{\half}{\frac{1}{2}}

\newcommand{\JJ}{\mathds{J}}
\newcommand{\KK}{\mathds{K}}

\newcommand{\DD}{\mathds{D}}
\newcommand{\TT}{\mathds{T}}
\newcommand{\ZZ}{\mathds{Z}}

\newcommand{\AAA}{\mathds{A}}
\newcommand{\FF}{\mathds{F}}

\newcommand{\calR}{\mathcal{R}}

\newcommand{\diff}[1]{\text{d}#1}
\newcommand{\Diff}[1]{\text{D}#1}

\newcommand{\Lag}{\mathscr{L}}

\newcommand{\bc}{\textcolor{blue}}

\newcommand{\uantof}{Departamento de Física, Universidad de Antofagasta, Aptdo. 02800, Chile}

\newcommand{\cecs}{Centro de Estudios Cient\'{\i}ficos (CECs), Arturo Prat 514, Valdivia, Chile}

\newcommand{\icen}{Instituto de Ciencias Exactas y Naturales, Universidad Arturo Prat, Playa Brava 3256, 1111346, Iquique, Chile}

\newcommand{\fdec}{Facultad de Ciencias, Universidad Arturo Prat, Avenida Arturo Prat Chac\'on 2120, 1110939, Iquique, Chile}

\newcommand{\uss}{Universidad San Sebastián, General Lagos 1163, Valdivia, Chile}

\begin{document}

\title{A black hole solution in conformal supergravity}

\author{Pedro D. Alvarez}
\email{pedro.alvarez@uantof.cl}
\affiliation{\uantof}

\author{Crist\'obal Corral}
\email{crcorral@unap.cl}
\affiliation{\icen}
\affiliation{\fdec}

\author{Jorge Zanelli}
\email{jorge.zanelli@uss.cl}
\affiliation{\cecs,}
\affiliation{\uss}

\begin{abstract}
We present a three-parameter family of analytic black-hole solutions in the bosonic sector of a four-dimensional supersymmetric model with matter fields in the adjoint representation. The solutions are endowed with a curvature and torsional singularities which are both surrounded by an event horizon. They are asymptotically Lorentz flat, representing the torsional generalization of the Riegert black hole in conformal gravity. We compute the partition function to first order in the saddle-point approximation which turns out to be finite without any reference to boundary counterterms. We find a non-maximmally symmetric thermalized ground state, whose existence is relevant when studying Hawking-Page phase transitions. Finally, we discuss future directions regarding its extended phase space.
\end{abstract}

\maketitle


\section{Introduction}\label{intro}

Supersymmetry (SUSY) is an appealing approach to address different problems in theoretical physics. From the viewpoint of quantum field theory, it attracted considerable interest as a mechanism to cancel divergences arising at the one-loop level coming from fermions and bosons. It is also a robust candidate for solving the hierarchy problem in the Standard Model of particle physics \cite{Martin:1997ns,Cohen:1996vb,Dimopoulos:1995mi}. Moreover, by combining spacetime and internal symmetries in a graded Lie algebra, SUSY circumvents the Coleman-Mandula theorem \cite{Coleman:1967ad}, offering a natural framework to unify those hitherto segregated symmetries. Supergravity, on the other hand, provides an interesting gravitational setup where the one-loop divergences are renormalized, in contrast to what happens in general relativity (for a review see~\cite{VanNieuwenhuizen:1981ae}). The former represents the gauge theory of SUSY and it appears as the low-energy limit of string theory~\cite{Blumenhagen:2013fgp,Ortin:2015hya}.


In recent years, a novel method of implementing SUSY, inspired by ideas from Yang-Mills theories, supergravity and Einstein-Cartan gravity, has been proposed~\cite{Alvarez:2011gd,Alv13,Alvarez:2015bva}. In this approach --dubbed unconventional supersymmetry (USUSY)--, the spin connection, the gauge potentials and matter fields belong to a gauge connection for a superextension of the anti-de Sitter (AdS) or Poincar\'e groups (for a review see~\cite{Alvarez:2021zhh}). What is unconventional about this approach is that matter fields are in the adjoint representation of the superalgebra. This feature is faithful to the spirit of supersymmetry, treating bosons and fermions on equal footing as parts of the same gauge connection. Fermionic fields representing matter are directly incorporated in the gauge connection by means of a Clifford algebra-valued soldering form that provides a metric structure and, eventually, the inclusion of gravity.

This way of implementing SUSY does not produce a pairing of boson-fermion states as in standard SUSY~\cite{Sohnius:1985qm,Alvarez:2011gd,Alv13} and therefore, USUSY models greatly differ from standard SUSYs and supergravities~\cite{Stelle:1978ye}.  The lack of superpartners remarkably coincides with the absence of supporting evidence for superpartners in the LHC. Recently, we have constructed chiral gauge models for $SU(2,2|2)$~\cite{Alvarez:2020qmy} and $SU(2,2|N)$~\cite{Alvarez:2021zsw}, and we have also described the embedding of the bi-fundamental spinor representation in the superconformal algebras~\cite{Alvarez:2021lda}. These developments can be seen as a step towards defining a grand unified theory based on the conformal superalgebras.

Conformal supergravity is an interesting gauge theory for the conformal superalgebra studied by many authors in the last decades~\cite{Kaku:1977pa,Ferrara:1977ij,Kaku:1978ea,Kaku:1978nz,Bergshoeff:1980is,Bergshoeff:1982az,Fradkin:1985am,Fradkin:1985am,Liu:1998bu,Butter:2009cp,Ferrara:2018wqd,DAuria:2021dth}. Indeed, by a proper gauge fixing, it is possible to arrive from conformal supergravity to the standard $\mathcal{N}=1$ supergravity in four dimensions~\cite{Freedman:2012zz}. In Refs.~\cite{Alvarez:2013tga,Alvarez:2020qmy, Alvarez:2021qbu}, we discussed a conformal supergravity theory based on the $\mathfrak{su}(2,2|N)$ gauge algebra written in a Townsend-MacDowell Mansouri~\cite{MacDowell:1977jt,Townsend:1977xw} form, which can be regarded as the $3+1$ generalization of the proposal in Ref.~\cite{Horne:1988jf}.\footnote{For a similar approach see Ref.~\cite{Andrianopoli:2014aqa}, whose applications in holography have been studied in Ref.~\cite{Andrianopoli:2020zbl}.} Its novelty relies on the implementation of supersymmetry in the adjoint representation, whose partially broken conformal symmetry is realized explicitly in the action. Remarkably, Dirac supercharges renders the full $R$-symmetry manifest~\cite{Trigiante:2016mnt}, where the latter is identified with the internal $SU(N)\times U(1)$ group. 

In the present paper, we explore some gravitational vacuum solutions of the theory, where ``vacuum"  means absence of fermionic matter fields. Similar configurations have been found in Chern-Simons AdS$_5$ supergravity models~\cite{Giribet:2014hpa,Andrianopoli:2021qli}. Here, we focus on four-dimensional topological black holes with constant-curvature transverse sections possessing torsional hair in unconventional conformal supergravity. We found a three-parameter family of analytic asymptotically locally Lorentz-flat solutions that represents the torsional generalization of the Riegert metric~\cite{Riegert:1984zz} in conformal gravity. We compute its temperature and partition function to first order in the saddle-point approximation. In a certain limit, we demonstrate that the black hole becomes a particular two-parameter family of solutions, possessing a central singularity dressed by a horizon, a nonvanishing temperature, but its free energy, mass, and entropy are identically zero. This class of spacetimes has been found in Weyl gravity as well and it has been regarded as a thermalized ground state~\cite{Lu:2012xu}. Here, we show their existence in unconventional conformal supergravity as well. 

The paper is organized as follows: In Sec.~\ref{model}, we define the basics of the model, namely, the field content, action principle, and field equations; essential ingredients for searching new vacuum solutions. In Sec.~\ref{solutions}, we present the novel topological black hole solution with torsion, alongside their properties. In Sec.~\ref{features}, we find the thermalized ground state as a limiting case of the three-parameter family of black holes. Finally, in Sec.~\ref{summary} we summarize our results.

\section{The model}\label{model}

Since we are interested in studying of vacuum solutions, we consider the purely bosonic sector of the theory discussed in~\cite{Alvarez:2021qbu}.\footnote{A general construction of this theory including spinors can be found in Ref.~\cite{Alvarez:2021qbu}.} The bosonic part of the gauge connection $1$-form is then given by
\begin{equation}\label{gaugeconnection}
\AAA= \half \omega^{ab}\JJ_{ab}+f^a \JJ_a+g^a\KK_a+h\DD+A^I \TT_I+A\ZZ \,,
\end{equation}
where $\JJ_{a}, \KK_{a}, \JJ_{ab}$, and $\DD$ denote the generators of the conformal group $SO(3,2)$, while $\TT_I$ and $\ZZ$ represent the generators of the internal $SU(N)$ and $U(1)$ groups, respectively. Hereon, upper and lowercase Latin indices label internal $SU(N)$ and Lorentz generators, respectively, while Greek indices denote spacetime coordinates. Each of the generators is contracted with its respective $1$-form compensating field and the field strength associated to this gauge connection is 
\begin{equation}
 \FF = \diff{\AAA} +\AAA \wedge \AAA =\half \calR^{ab} \JJ_{ab}+F^a \JJ_a +G^a \KK_a +H \DD +F^I \TT_I +F \ZZ \,,
\end{equation}
where $\diff{}$ is the exterior derivative, $\wedge$ is the exterior product of differential forms. The explicit expressions of the curvature components are 
\begin{align}
\mathcal{R}^{ab}&=R^{ab}+f^a\wedge f^b-g^a\wedge g^b\,, & F^a&=\Diff{f}^a+g^a\wedge h\,, & G^a&=\Diff{g}^a+f^a\wedge h\,,\\
H  &=\diff{h}+f^a\wedge g_a\,, & F^I&=\diff{A}^I+\frac{1}{2} f_{JK}^{I}A^J\wedge A^K\,, & 
F &=\diff{A}\,,
\end{align}
with $R^{ab}=\diff{\omega^{ab}}+\omega^a_{\ c}\wedge\omega^{cb}$ being the Lorentz curvature $2$-form and $\Diff{}$ denotes the covariant derivative with respect to the Lorentz connection. 

The dynamics of the theory is described by a generalization of the MacDowell-Mansouri action~\cite{MacDowell:1977jt} for the bosonic sector of the superconformal group, that is,
\begin{equation}\label{action}
 \mathcal{S}= - \int \langle \FF \wedge \circledast\ \FF \rangle\,,
\end{equation}
where $\langle\ldots\rangle$ denotes trace over the internal indices and the dual operator $\circledast$ is defined through
\begin{equation}
\circledast \FF= S\left(\half \calR^{ab} \JJ_{ab}+F^a \JJ_a +G^a \KK_a\right)+(\varepsilon_1\ast)H \DD +(\varepsilon_2\ast) F^I \TT_I +(\varepsilon_3\ast)F \ZZ \,. \label{dualop}
\end{equation}
Here $\varepsilon_i=\pm1$, with $i=1,2,3$, $\ast$ is the Hodge dual, and $S=i\gamma_5$ with $\gamma_5$ being the chiral gamma matrix defined such that $S^2=-\mathbb{I}$. The ambiguity in $\varepsilon_i$ can be eliminated by demanding the correct sign of the kinetic terms of the bosonic fields in the action. This, in turn, will depend on the details of the algebra representation. More details of this operator can be found in~\cite{Alvarez:2021zsw}.

The action functional~\eqref{action} can be thought of as a Yang-Mills theory for an embedding $G^+ \hookrightarrow SU(2,2|N)$. Using the properties of the supertrace (cf. Ref.~\cite{Alvarez:2021zsw}), we can express the the bosonic sector of the Lagrangian along the curvature components as
\begin{align}\notag
\Lag_\text{bos}&=\frac{1}{4}\epsilon_{abcd}\calR^{ab}\wedge\calR^{cd}-\varepsilon_1 (H+f^a\wedge g_a)\wedge\ast (H+f^b\wedge g_b) \\
&-\half\varepsilon_2 F^I\wedge\ast F^I -4\left(\frac{4}{N}-1\right) \varepsilon_3 F\wedge\ast F\,.\label{actionexpandbos}
\end{align}
A ghost-free action demands $\varepsilon_1=\varepsilon_2=1$. The value of $\varepsilon_3$, on the other hand, depends on $N$: the absence of ghosts demands $\varepsilon_3=\pm1$ for $N<4$ and $N>4$, respectively, such that $\varepsilon_3(N-4)<0$. For $N=4$, however, the $U(1)$ sector drops out from the Lagrangian~\eqref{actionexpandbos}.

Varying the Lagrangian~\eqref{actionexpandbos} with respect to $f^a$ and $g^a$ yields
\begin{align}
 \epsilon_{abcd}f^b\wedge\calR^{cd}-2 g_a\wedge\ast(H+f^b\wedge g_b) &= 0\,,\label{eqfa}\\
 \epsilon_{abcd}g^b\wedge\calR^{cd}-2 f_a\wedge\ast(H+f^b\wedge g_b) &= 0\,,\label{eqga}
\end{align}
respectively. We will search for solutions of these equations~\eqref{eqfa} and~\eqref{eqga} under the assumption of Lorentz invariance, which is naturally guaranteed if both $f^a$ and $g^a$ are conformally related to a soldering form $e^a$ that provides the metric structure for the theory. This means $f^a = \rho(x) e^a$, $g^a = \sigma(x) e^a$, where $\rho(x)$ and $\sigma(x)$ are arbitrary scalar functions and $e^a = e^a_\mu\diff{x^\mu}$ with $g_{\mu \nu}= \eta_{ab}\, e^a_\mu\, e^b_\nu$. This choice breaks the conformal invariance and, therefore, the Weyl rescaling is no longer a gauge symmetry. Then, in the simplest version of this theory, we can take $h=0$. These assumptions imply $\mathcal{R}^{ab} = R^{ab} - \Lambda(x) e^a\wedge e^b$, where we have defined $\Lambda(x):=\sigma^2(x)-\rho^2(x)$. Note that asymptotically locally AdS solutions are allowed provided $\Lambda(x)<0$. 

The remaining field equations are obtained by extremizing the Lagrangian~\eqref{actionexpandbos} with respect to the dynamical fields in the AdS sector. Then, the full set of field equations is
\begin{align}\label{eom2}
 \epsilon_{abcd} \Diff{} \calR^{cd}&=0\,,& \epsilon_{abcd}e^b\wedge \calR^{cd}&=0\,, &  \mathfrak{D}\ast F^I&=0\,, & \diff{} \ast F &=0\,,
\end{align}
where $\mathfrak{D}\equiv D_{(A^J \TT_J)}$ is the covariant derivative for the internal gauge connection. These equations, alongside~\eqref{eqfa} and~\eqref{eqga}, are second-order dynamics equations for the gauge connection $\AAA$ that determine the states of the theory. For the sake of simplicity, we henceforth focus in the sector with vanishing $SU(N)$ and $U(1)$ gauge fields. 

In order to look for the ground state of the theory, we notice that these equations admit $\calR^{ab}=0$ as solution. Choosing $\rho(x)^2-\sigma(x)^2=\pm\ell^{-2}$ to be constant, locally AdS or dS spacetimes are obtained, that is,
\begin{equation} \label{(A)dS}
 R^{ab} \pm \ell^{-2}e^a\wedge e^b=0\,.
\end{equation}
These geometries include maximally symmetric spaces with up to $10$ globally defined Killing vectors, which can be genuinely interpreted as vacuum configurations. The symmetric case $\rho(x)^2=\sigma(x)^2$, on the other hand, yields
\begin{equation}\label{flat}
\epsilon_{abcd}e^b\wedge R^{cd} =0\,,
\end{equation}
which includes locally Lorentz-flat configurations, $R^{ab}=0$, as well as nontrivial asymptotically locally Lorentz-flat solutions. This can be achieved if the torsion $2$-form is convariantly conserved. We present a solution of the former class in the next section.

\section{Static spherically symmetric solution}\label{solutions}

Here, we look for black hole solutions to the field equations~\eqref{eom2}. We work in the first-order formalism where the vielbein $1$-form, $e^a=e^a{}_\mu\diff{x^\mu}$, and the Lorentz connection $1$-form, $\omega^{ab}=\omega^{ab}{}_\mu\diff{x^\mu}$, define the Lorentz curvature and torsion $2$-forms through the Cartan structure equations as 
\begin{align}\label{RandT}
R^{ab}&=\diff{\omega^{ab}} + \omega^{a}{}_c\wedge\omega^{cb}=\tfrac{1}{2}R^{ab}_{\ \mu\nu}\diff{x^\mu}\wedge\diff{x^\nu}\,, \\
T^a &= \diff{e^a} + \omega^{a}{}_b\wedge e^b=\tfrac{1}{2}T^{a}_{\ \mu\nu}\diff{x^\mu}\wedge\diff{x^\nu}\,,    
\end{align}
respectively. These quantities satisfy the Bianchi identities $\Diff{}R^{ab}=0$ and $\Diff{} T^a = R^{a}{}_b\wedge e^b$. The Lorentz connection $1$-form can be decomposed as $\omega^{ab}=\mathring{\omega}^{ab} + \kappa^{ab}$, where $\mathring{\omega}^{ab}$ denotes its torsion-free part satisfying $\diff{e^a} + \mathring{\omega}^{a}{}_b\wedge e^b=0$ and $\kappa^{ab}$ is the contorsion $1$-form defined through $T^a = \kappa^{a}{}_b\wedge e^b$. This decomposition allows rewriting the Lorentz curvature $2$-form as $R^{ab}=\mathring{R}^{ab} + \mathring{\Diff{}}\kappa^{ab} + \kappa^{a}{}_c\wedge\kappa^{cb}$, where $\mathring{R}^{ab}=\tfrac{1}{2}\mathring{R}^{ab}_{\ \mu\nu}\diff{x^\mu}\wedge\diff{x^\nu}$ represents its torsion-free part related to the Riemann tensor through $\mathring{R}^{\lambda\rho}_{\ \mu\nu} = e^\lambda{}_a e^\rho{}_b\mathring{R}^{ab}_{\ \mu\nu}$ with $e^\mu{}_a$ being the inverse vielbein, while $\mathring{\Diff{}}\kappa^{ab}$ denotes the covariant derivative of the contorsion $2$-form with respect to the torsion-free Lorentz connection. From hereon, ringed quantities denote torsion-free geometric objects constructed out of $\mathring{\omega}^{ab}$. 

In order to search for static topological black-hole solutions to the field equations~\eqref{eom2}, we focus on a metric ansatz possessing a constant-curvature base manifold, namely,
\begin{equation}
\label{metricansatz}
\diff{s^2}= g_{\mu\nu}\diff{x^\mu}\otimes\diff{x^\nu} = -f(r)\diff{t^2}+\frac{\diff{r^2}}{f(r)}+r^2\text{d}\Sigma^2_{(k)}\,,
\end{equation}
where the line element of the two-dimensional base manifold is parametrized as
\begin{align}\label{basek}
 \diff{\Sigma_{(k)}^2} = \bar{g}_{\bar{\mu}\bar{\nu}}\diff{\bar{x}^{\bar{\mu}}}\otimes\diff{\bar{x}^{\bar{\nu}}} = \delta_{\bar{a}\bar{b}}\bar{e}^{\bar{a}}\otimes \bar{e}^{\bar{b}}\,.
\end{align}
Here, barred quantities are intrinsically defined on the two-dimensional transverse section $\diff{\Sigma_{(k)}^2}$ and $k=\pm1,0$ stands for spherical, hyperbolic or flat topology, respectively. According to the relation $g_{\mu\nu} = \eta_{ab}e^a_\mu e^b_\mu$, we choose a vielbein basis 
\begin{align}\label{eansatz}
e^0&=\sqrt{f(r)}\,\diff{t}\,, &  e^1&=\frac{\diff{r}}{\sqrt{f(r)}}\,, &  e^{\bar{a}}&=r\,\bar{e}^{\bar{a}}\,.
\end{align}
Demanding invariance under the same isometry group of the metric~\eqref{metricansatz} to the remaining bosonic fields,\footnote{When $k=\pm1,0$, the metric~\eqref{metricansatz} is locally invariant under the action of $\mbox{SO}(3)\times\mathbb{R}$, $\mbox{SO}(1,2)\times\mathbb{R}$, or $\mbox{ISO}(2)\times\mathbb{R}$ isometry groups, respectively. To obtain the ansatz~\eqref{omegaansatz}, we decompose $\omega^{ab}=\mathring{\omega}^{ab}+\kappa^{ab}$ and we read off their independent components from the invariance of $\mathring{\omega}^{ab}$ and $\kappa^{ab}$ under these isometry groups.} we find $\Lambda=\Lambda(r)$, and 
\begin{subequations}\label{omegaansatz}
\begin{align}
 \omega^{01} &= \omega_1(r) e^0 + \omega_2(r) e^1\,, & \omega^{02} &= \omega_3(r) e^2 + \omega_4(r) e^3\,, \\
 \omega^{03} &= -\omega_4(r) e^2 + \omega_3(r) e^3\,, & \omega^{12} &= \omega_5(r) e^2 + \omega_6(r) e^3\,, \\
 \omega^{13} &= -\omega_6(r) e^2 + \omega_5(r) e^3\,, &
 \omega^{23} &= \omega_7(r) e^0 + \omega_8(r) e^1 + \mathring{\omega}^{23}\,.
\end{align}
\end{subequations}
In this case, $\omega_i(r)$, with $i=1,\dots,8$, are indeterminate functions depending on the radial coordinate only. The nontrivial components of the Lorentz curvature and torsion can be easily found from Eq.~\eqref{RandT} alongside Eqs.~\eqref{eansatz} and~\eqref{omegaansatz}. 

Inserting these ans\"atze into the field equations~\eqref{eom2}, we find 
\begin{subequations}\label{odesys}
\begin{align}
    \omega_2=\omega_3=\omega_4=\omega_6=\omega_7=\omega_8&=0\,, \\
    \Lambda' f r + 2f\Lambda + 2\sqrt{f}\,r\Lambda\omega_5 &=0 \,, \\
    \left(f r^2 \Lambda^2 \right)' + 2\sqrt{f}\,r^2\Lambda^2\left(\omega_5-\omega_1\right) &= 0\,, \\
    2\omega_5'f\,r^2 + 2f\omega_5\,r + k\sqrt{f} + 3\sqrt{f}\,\Lambda\,r^2-\sqrt{f}\omega_5^2\,r^2&=0\,, \\
    \frac{k}{r^2} + 3\Lambda + 2\omega_1\omega_5 - \omega_5^2 &= 0\,, 
\end{align}
\end{subequations}
where prime denotes differentiation with respect to $r$. This nonlinear system of first-order ordinary differential equations is solved by
\begin{subequations}\label{lasolucion}
\begin{align}\label{solutionskarb}
    f(r) &= \gamma - \frac{2m}{r} - \frac{(\gamma^2-k^2)r}{6m} + a\,r^2  \,, & \omega_5(r) &= \frac{\sqrt{f}(k+\frac{1}{2}f''r^2-f'r+f)}{r\left(f'r-2f \right)}\,, \\
    \omega_1(r) &= \sqrt{f}\,\frac{\diff{}}{\diff{r}}\ln (\omega_5\, r)\,, & \Lambda(r) &=  \frac{k-\omega_5\,r^2\left(\omega_5-2\omega_1\right)}{3r^2}\,.\label{solutionskarb2}
\end{align}
\end{subequations}
This solution represents a three-parameter family of topological black holes characterized by the integration constants $\gamma$, $m$, and $a$. It can be checked that, as $r\to \infty$, the metric function behaves $f \sim ar^2+\mathcal{O}(r)$, and therefore $\omega_1 \sim \omega_5 \to 0$. Then, from Eq.~\eqref{solutionskarb2}, one finds that $\Lambda \to 0$. Thus, even though the solution represents a weakened asymptotically AdS metric, the geometry is asymptotically Lorentz flat from a Riemann-Cartan viewpoint. A similar feature has been observed for locally AdS configurations in three dimensions that represents Lorentz-flat spacetimes in presence of covariantly constant torsion~\cite{Alvarez:2014uda}. Indeed, the solution~\eqref{lasolucion} resembles this behavior asymptotically in four dimensions. The nonvanishing components of the Lorentz curvature and torsion are given by
\begin{subequations}\label{R&Tformsk}
\begin{align}
    R^{01} &= R_I(r)\, e^0\wedge e^1\,, & R^{0\bar{a}} &= R_{II}(r)\,e^0\wedge e^{\bar{a}}\,, & R^{1\bar{a}} &= R_{II}(r)\, e^1\wedge e^{\bar{a}}\,, \\
    R^{23} &= R_I(r)\,e^2\wedge e^3\,, & T^a &= T_I(r)\, e^1\wedge e^a\,, & T^{\bar{a}} &=T_{II}(r)\, \epsilon^{\bar{a}}{}_{\,\bar{b}\bar{c}}\, e^{\bar{b}} \wedge e^{\bar{c}} \;, 
\end{align}
\end{subequations}
where the particular values for the solution~\eqref{lasolucion} are
\begin{subequations}\label{R&Tcompk}
\begin{align}
 R_I(r) &= \frac{\left[k\left(\gamma-k \right)^2-36am^2 \right]r^3 + 6m(\gamma-k)^2r^2 - 36m^2(\gamma-k)r+72m^3}{\left[(\gamma-k)r-6m \right]^2r^3} \,, \\
 R_{II}(r) &= \frac{\left[(\gamma+k)(\gamma-k)^2-72am^2 \right]r^3 - 6m(\gamma-k)^2r^2+36m^2(\gamma-k)r-72m^3}{2\left[(\gamma-k)r-6m\right]^2r^3} \,,\\
 \label{TIkarb}
 T_I(r) &= \frac{\gamma-k}{(\gamma-k) r - 6m} \sqrt{ ar^2-\frac{(\gamma^2-k^2)}{6m}r + \gamma - \frac{2m}{r}}\,,\\
 T_{II}(r) &=0\,.
\end{align}
\end{subequations}

We shall discuss the main properties of this three-parameter family of black-hole solutions in the next section.

\section{Features of the torsional black hole}\label{features}

Here, we analyze different aspects of the solution including the structure of singularities and their thermodynamic properties in presence of torsion. Moreover, we derive a particular limit where it becomes a thermalized ground state.

\subsection{Curvature}

The linear term in $r$ of the metric function $f(r)$ sources the non-Einstein mode of the solution. This can be seen as follows: First, consider the torsion-free Riemann tensor $\mathring{R}^{\mu\nu}_{\ 
\lambda\rho}$ constructed out of the Levi-Civita connection associated to the line element~\eqref{metricansatz} with metric function~\eqref{solutionskarb}.  In general, the traceless Ricci tensor $\mathring{H}_{\mu\nu} \equiv \mathring{R}_{\mu\nu} - \frac{1}{4}g_{\mu\nu}\mathring{R}$ vanishes identically for Einstein spaces. The curvature invariant 
\begin{align}\label{Hcuadrado}
    \mathring{H}_{\mu\nu}\mathring{H}^{\mu\nu} = \frac{(\gamma-k)^2[(\gamma+k)r-6m]^2}{36m^2r^4}\,,
\end{align}
shows that, in this case, $\mathring{H}_{\mu\nu}$ is nonvanishing. Thus, we conclude that Eq.~\eqref{lasolucion} represents a non-Einstein space. It is worth mentioning that Maldacena showed in Ref.~\cite{Maldacena:2011mk} that imposing Neumann boundary conditions on the Fefferman-Graham expansion of the weakened asymptotically locally AdS metric allows one to remove the linear mode present in conformal gravity. This procedure selects Einstein spaces from the space of solutions of conformal gravity, whose renormalization is guaranteed for asymptotically locally AdS spaces due to conformal invariance~\cite{Grumiller:2013mxa}.    

The solution is Bach-flat, as it can be checked by noticing that the Bach tensor,
\begin{align}
    \mathring{B}_{\mu\nu} = \nabla^\lambda \mathring{C}_{\mu\nu\lambda} - \mathring{S}^{\lambda\rho}\mathring{W}_{\mu\lambda\rho\nu}\,,
\end{align}
vanishes identically for the solution~\eqref{lasolucion}, where $\mathring{C}_{\mu\nu\lambda}=2\nabla_{[\lambda}\mathring{S}_{\nu]\mu}$ and $\mathring{S}_{\mu\nu}=\tfrac{1}{2}\left(\mathring{R}_{\mu\nu}-\tfrac{1}{6}g_{\mu\nu}\mathring{R} \right)$ are the torsion-free Cotton and Schouten tensor, respectively. Thus, we conclude that spacetime \eqref{lasolucion} can be regarded as a torsional generalization of the Riegert metric found in the second-order (purely metric) formulation of conformal gravity \cite{Riegert:1984zz}. Notice that, in the presence of the extra fields required by USUSY, torsion is switched on, so that the first and second order formulations are not equivalent: the torsional features of the geometry do not vanish. These torsional contributions can be moved to the right hand side of the field equations~\eqref{eom2}, acting as a source for the torsion-free part of the geometry. This occurs in many situations in which the first and second order formalisms are not equivalent~\cite{Toloza:2013wi,Espiro:2014uda,Castillo-Felisola:2015ema,Castillo-Felisola:2016kpe,Barrientos:2017utp,Cid:2017wtf,Cisterna:2018jsx,Barrientos:2019awg}.

For $\gamma\neq k$, the solution~\eqref{lasolucion} has a curvature singularity at $r=0$ as it can be seen from the torsion-free Ricci scalar [see also Eq.~\eqref{Hcuadrado}]
\begin{align}\label{Ricciscalfsol}
    \mathring{R} &= \mathring{R}^{\mu\nu}_{\ \mu\nu} = -12a + \frac{\left(\gamma^2-k^2\right)}{m\,r} - \frac{2(\gamma-k)}{r^2}\,.
\end{align}
The central singularity is surrounded by a horizon located at $r=r_h$, defined by the condition $f(r_h) = 0$. This is a cubic equation for $r_h$ having one real root and two complex conjugate ones. The real one represents the black hole's horizon whose explicit form is rather cumbersome and not very illuminating. The conditions on the integration constants that make $r_h>0$ are discussed in Eq.~\eqref{condparahorizonte} below. 

\subsection{Torsion}
Besides the singularity at $r=0$, there exists a torsional singularity at $r_s  \equiv 6m(\gamma-k)^{-1}$, as can be seen from the invariant
\begin{align}
T_{\mu\nu\lambda}T^{\mu\nu\lambda} 
&= \frac{6}{r_s(r-r_s)^2}\,\left[a r_s\,r^2 - (\gamma+ k)r +\gamma \,r_s -\frac{2m r_s}{r}\right]   \,,
\end{align}
where $T^{\lambda}_{\ \mu\nu} = e^\lambda{}_a T^{a}_{\ \mu\nu}$. The conditions on the parameters of the solution that ensure that the torsional singularity lies inside the black hole's event horizon, $r_h>r_s$, are discussed below.

From \eqref{R&Tcompk} it is apparent that the Lorentz curvature $R^{ab}$ and the torsion components $T^a$ diverge both at $r=r_s$ and at $r=0$. In fact, it is also the case for all polynomial invariants obtained by contracting these tensors. However, it is a remarkable feature of the solution that torsion-free quantities, like the Ricci scalar \eqref{Ricciscalfsol}, are regular at $r=r_s$. This is true of other torsion-free geometric quantities, like the Kretchsmann or the quadratic curvature invariant \eqref{Hcuadrado}. In particular, the torsion-free Weyl tensor
$\mathring{W}^{\mu\nu}_{\lambda\rho}= \mathring{R}^{\mu\nu}_{\lambda\rho} - 4\delta^{[\mu}_{[\lambda}\mathring{S}^{\nu]}_{\rho]}$,
\begin{align} \label{weyl}
    \mathring{W}^{\mu\nu}_{\lambda\rho} &= - \frac{(\gamma-k)r-6m}{3r^2}\delta^{\mu\nu}_{\lambda\rho}\,,
\end{align}
is regular everywhere except at $r=0$. In fact, this solution is conformally flat at $r=r_s$ $\mathring{W}^{\mu\nu}_{\lambda\rho}$ where \eqref{weyl} vanishes.

The torsional singularity does not affect the motion of spinless test particles since they couple to the torsion-free Christoffel connection and follow geodesics rather than autoparallels~\cite{Hehl:1976kj,Hehl:1994ue}. Spinning particles and fermions in general, on the other hand, couple to torsion \cite{Hehl:1971qi,Hehl:1976vr,Chandia:1998nu}. Therefore, their propagation must be sensitive to the torsional singularity in this background. This should be taken into account when studying the geodesic completeness of this solution for spinning test particles.          

The existence of a horizon plus the condition $r_s<r_h$, that enforces the cosmic censorship principle, provides a restriction on the parameters. For the sake of simplicity, consider the particular case of weakened asymptotically AdS solution with $m>0$, $a>0$, $k=1$, and $-1\leq\gamma\leq1$. In this case, the conditions $0<r_s<r_h$ translate into 
\begin{align}\label{condparahorizonte}
 -1\leq\gamma<1  \;\;\;\;\; \wedge \;\;\;\;\; \frac{2-3\gamma+\gamma^3}{108m^2}\leq a \;\;\;\;\; \vee \;\;\;\;\; \gamma=1\;\;\;\;\; \wedge\;\;\;\;\; 0<a\,.
\end{align}
Similar restrictions on parameter space are found for the asymptotically de Sitter case so that the curvature and torsional singularities are hidden by the event horizon. Nevertheless, if $a<0$, the solution is endowed with a cosmological horizon as well.

The solution~\eqref{lasolucion} possesses a covariantly constant torsion, which can be checked by using the Bianchi identity $\Diff{T^a}=R^{a}{}_b\wedge e^b$ and noticing that the Lorentz curvature $2$-form~\eqref{R&Tcompk} satisfies $R^{a}{}_b\wedge e^b=0$. Geometries with covariantly constant torsion include constant Lorentz curvature solutions and Lorentz-flat geometries as particular cases. In three dimensions, the most general solution to $\Diff{}T^a=0$ has at least one pseudoscalar degree of freedom~\cite{Alvarez:2014uda} and admits a Proca-like excitation~\cite{Andrianopoli+}. In four or higher dimensions additional degrees of freedom could be present. 

The torsional Riegert black hole~\eqref{lasolucion} is continuously connected to the torsion-free Schwarzschild-AdS and dS with constant curvature base manifold in the limit $\gamma\to k$ for $a>0$ and $a<0$, respectively. When $\gamma=-k$, however, the linear mode in the metric~\eqref{solutionskarb} is absent and the torsion is nonvanishing. Nevertheless, this case represents a naked singularity located at $r=0$, since there is no set of parameters that allows for a horizon. When $\gamma=k=0$, the metric becomes the cylindrical black hole studied by Lemos in Ref.~\cite{Lemos:1994xp}. Thus, a similar large gauge transformation could give rise to a nonzero angular momentum when $k=0$ and $\gamma\neq0$ such that the torsion is nonvanishing. However, in this work we focus on the static case for the sake of simplicity and we postpone stationary extensions for future studies. 

\subsection{Thermodynamics}

The thermodynamic properties of these black holes can be derived from the partition function $\mathcal{Z}$ (see~\cite{Gibbons:1976ue,Hawking:1982dh} for details). This can be computed to first order in the saddle-point approximation through $\ln\mathcal{Z}\approx -I_E$, where $I_E$ is the Euclidean on-shell action. The latter can be obtained by performing the analytic continuation $t\to -i\tau$ in the line element~\eqref{metricansatz}. The absence of conical singularities when the Euclidean time coordinate is identified as $\tau\sim\tau+\beta$ fixes thiss period to be $\beta=4\pi/f'(r_h)$. Then, the Hawking temperature for the torsional Riegert black hole given  by~\eqref{lasolucion} is
\begin{align}
    T_H 
    = \frac{\gamma+k}{4\pi r_s} +\frac{3m-\gamma r_h}{2\pi r_h^2}\,.
\end{align} 
In contrast to the Schwarzschild-AdS solution, the temperature of the torsional Riegert black hole behaves as
\begin{align}
    T_H \approx \frac{\gamma^2-k^2}{24\pi m} + \mathcal{O}(r_h^{-1}) \;\;\;\; \mbox{when} \;\;\;\; r_h\to\infty\,.
\end{align}
In the torsion-free limit $\gamma\to k$, however, it behaves as usual, namely, $T_H\approx\mathcal{O}(r_h^{-1})$ as $r_h\to\infty$. 

A direct evaluation of the Euclidean on-shell action~\eqref{action} on the three-parameter family of black-hole solutions~\eqref{lasolucion} gives the partition function to first-order in the saddle point approximation, that is,
\begin{align}\label{IE}
    \ln\mathcal{Z} \approx -I_E = \frac{\beta\Omega_{(k)}\left[(\gamma-k)^2 r_h^2 - 6m(\gamma-k)r_h + 12m^2 \right]}{3r_h^3}\,,
\end{align}
where $\Omega_{(k)}$ denotes the volume of the base manifold. Remarkably, conformal invariance renders its value finite without any reference to boundary counterterms. This feature has been observed in different theories possessing conformal invariance in the metric formalism~\cite{Grumiller:2013mxa,Anastasiou:2020mik,Anastasiou:2021tlv,Barrientos:2022yoz} and we hereby provide additional evidence for conformal supergravity with nontrivial torsion. Moreover, the cosmological term in the solution~\eqref{lasolucion} arises as an integration constant, offering a natural scenario to study the extended phase space of black hole thermodynamics as proposed in~\cite{Kubiznak:2016qmn}, as well as their critical behavior~\cite{Kubiznak:2012wp}. Due to the number of free parameters, the phase structure will be richer and we postpone a deeper study of this point for the future.

\subsection{Thermalized ground state}\label{thermalground}

An interesting limit of the solution~\eqref{lasolucion} is $\gamma\to k$ and $m\to0$, with $k(\gamma-k)(3m)^{-1} \equiv b$ (or $\frac{\gamma-k}{6m}=r_s^{-1}$) fixed. In this limit, the torsional singularity is translated into $r_s=2kb^{-1}$. Then, we obtain 
\begin{align}\label{solution}
    f(r) = k - br + ar^2\,,
\end{align}
where $\omega_1(r)$, $\omega_5(r)$, and $\Lambda(r)$ can be obtained directly from the relations in Eq.~\eqref{lasolucion}. This solution possesses a weakened AdS or dS asymptotics when $a>0$ or $a<0$, respectively, and it is weakly asymptotically locally flat when $a=0$. In this case, the Lorentz curvature and torsion $2$-forms are
\begin{align}\label{R&Tcomp}
    R^{ab} 
    &= \frac{k-a\,r_s^2}{\left(r-r_s \right)^2} \,e^a\wedge e^b \;\;\;\;\; \mbox{and} \;\;\;\;\;  T^a = \frac{\sqrt{f(r)}}{r-r_s}\,e^1\wedge e^a \,.
\end{align}
Additionally, one can check that this spacetime is both conformally and Bach flat. 

From the Bianchi identity $\Diff{T^a}=R^{a}{}_b\wedge e^b$, it is direct to see that the torsion $2$-form is covariantly conserved as well, namely, $DT^a=0$. Moreover, this solution has curvature and torsional singularities at $r=0$ and $\bc{r=r_s}$, respectively. The horizons are located at the two positive real roots of the metric function $f(r)$, i.e. $f(r_\pm) = 0$, where 
\begin{align}
    r_\pm = \frac{1}{2a}\left(b\pm\sqrt{b^2-4ak} \right)\,.
\end{align}

The existence of a horizon requires $b^2-4ak\geq0$, and the bound is saturated in the extremal case, which can only happen if $ak>0$. In the non-extremal case, the condition $ak<0$ guarantees that a horizon will always exist. Additionally, demand the torsional singularity to lie behind the horizon by imposing $r_+>r_s$. Note that the torsion vanishes at the horizon, as it can be seen from Eq.~\eqref{R&Tcomp}. Moreover, in the extremal case, there is no torsional singularity whatsoever, as it can be checked from the torsional invariants that become constant, e.g. $T_{\mu\nu\lambda}T^{\mu\nu\lambda} = \tfrac{3b^2}{2k}$. Notice that the solution~\eqref{solution} is non-Einstein for $b\neq0$, as can be seen from the torsion-free traceless Ricci tensor $\mathring{H}$, that is, 
\begin{align}
    \mathring{H}_{\mu\nu}\mathring{H}^{\mu\nu} = \frac{b^2}{r^2} \,.
\end{align}

The solution~\eqref{solution} has a rather peculiar thermodynamic behavior. First, its Hawking temperature is nonvanishing and is given by 
\begin{align}\label{THgroundstate}
    T_H 
    = \frac{k}{2\pi r_+\,r_s}(r_+ - r_s)\,. 
\end{align}
The requirement that the torsional singularity must lie behind the black hole's horizon, i.e. $r_+>r_s$, implies that the Hawking temperature is positive definite. However, a direct computation of the Euclidean on-shell action shows that it vanishes identically. This implies that its free energy, mass, and entropy are zero. Therefore, we conclude that this limit represents a conformally flat ground state of the theory that is not a maximally symmetric space, although continuously connected to global AdS or dS in the limit $b\to0$ for $a>0$ or $a<0$, respectively. Nevertheless, since its Hawking temperature~\eqref{THgroundstate} is nonvanishing, it can be regarded as a thermalized ground state, similar to the Schwarzschild-AdS solution in critical gravity~\cite{Lu:2011zk}. This ground state has been studied in metric formulation of conformal gravity~\cite{Lu:2012xu} and the solution~\eqref{solution}--\eqref{R&Tcomp} represents its torsional generalization embedded in the bosonic sector of unconventional conformal supergravity.   

\section{Summary}\label{summary}

We have explored a sector of the space of solutions of a particular formulation of conformal supergravity. First, we check that this theory possesses maximally symmetric spaces as ground states. Additionally, we show that black-hole solutions also exist; the latter can be regarded as the torsional extension of the Riegert spacetime~\cite{Riegert:1984zz}. The compensating fields for translations and special conformal transformations of the $SO(3,2)$ group generate a backreaction that behaves as a dynamical cosmological constant. Indeed, we find that the solution is a weakened asymptotically locally AdS, however, from the Riemann-Cartan viewpoint, it represents a novel example of an asymptotically Lorentz-flat black hole.      

This solution presents a number of interesting features. First, it possesses a torsional singularity where the metric becomes conformally flat. Spinning test particles should be sensitive to this kind of singularities, providing an interesting setup to study autoparallel completeness rather than geodesic completeness. The conditions on the integration constants of the solution are found such that the curvature and torsion singularities lie behind the horizon, avoiding a violation of the cosmic censorship conjecture. When $SU(N)$ and $U(1)$ connections are turned off, the black-hole solution is continuously connected to Schwarzschild-(A)dS in the torsion-free limit. Additionally, the torsion is covariantly conserved, something that could be relevant for computing conserved charges. In a particular limit, the torsional black hole becomes a thermalized ground state, namely, a non-maximally symmetric configuration with positive definite temperature whose free energy vanishes. This configuration is likely to be of relevance in the study of the realted Hawking-Page phase transitions.

Other interesting properties of this theory can be explored in the future. For instance, since the cosmological constant term in Eq.~\eqref{lasolucion} appears as an integration constant, this theory offers a natural scenario to study the extended phase space of the torsional Riegert black hole in the Gibbs canonical ensemble by interpreting the cosmological term as a thermodynamic pressure~\cite{Kubiznak:2016qmn}. On the other hand, stationary solutions in unconventional conformal supergravity are certainly of interest to us. We believe that the role of torsion will be important in that case due to the axial symmetry involved in this type of configurations and it is expected that novel topological features associated to torsion could arise (see~\cite{Chandia:1997hu}). Finally, holographic properties of torsion can be studied in conformal supergravity, since it is interpreted as the source of the spin density in the dual field theory \cite{Banados:2006fe,Klemm:2007yu,Blagojevic:2013bu}. We expect to return for a deeper study of these aspects in the near future.

\begin{acknowledgments}
We thank Giorgos Anastasiou, Laura Andrianopoli, Ignacio J. Araya, Oscar Castillo-Felisola, Rodrigo Olea, and Omar Valdivia for useful discussions. P.~A. acknowledges MINEDUC-UA project ANT 1755 and Semillero de Investigación project SEM18-02 from Universidad de Antofagasta, Chile. The work of C.~C. and J.~Z. is supported by Agencia Nacional de Investigación y Desarrollo (ANID) through FONDECYT No 11200025, 1210500, and 1220862.
\end{acknowledgments}

\bibliography{paper.bib}

\end{document}